\documentclass{article}
\usepackage{spconf,amsmath,graphicx,hyperref,booktabs}

\title{Towards effective negation modeling \\ in joint audio-text models for music}
\name{$^{1}$Yannis Vasilakis, $^{2}$Rachel Bittner, $^{1}$Johan Pauwels\thanks{The first author is a research student at the UKRI Centre for Doctoral Training in Artificial Intelligence and Music, supported jointly by UK Research and Innovation [grant number EP/S022694/1] and Queen Mary University of London.}} 
\address{$^{1}$Queen Mary University of London, $^{2}$Spotify}
\begin{document}
%
\maketitle
\begin{abstract}
Joint audio–text models are widely used for music retrieval, yet they struggle with semantic phenomena such as negation. Negation is fundamental for distinguishing the absence (or presence) of musical elements (e.g., “with vocals” vs. “without vocals”), but current systems fail to represent this reliably. In this work, we investigate and mitigate this limitation by training CLAP models from scratch on the Million Song Dataset with LP-MusicCaps-MSD captions. We introduce negation through text augmentation and a dissimilarity-based contrastive loss, designed to explicitly separate original and negated captions in the joint embedding space. To evaluate progress, we propose two protocols that frame negation modeling as retrieval and binary classification tasks. Experiments demonstrate that both methods, individually and combined, improve negation handling while largely preserving retrieval performance.
\end{abstract}
\begin{keywords}
multimodal, joint audio-text models, music, negation, retrieval, music understanding
\end{keywords}

\section{Introduction}
\label{sec:introduction}

Joint audio-text models have emerged as effective tools for text-to-audio and audio-to-text retrieval, providing a compact way to combine textual and audio information. These models have demonstrated success in retrieval tasks and music classification~\cite{muscall_manco_2022}, and have been used for text-conditioned audio~\cite{Liu2023AudioLDMTG} or music generation~\cite{DBLP:conf/ismir/NistalPAGL24}.

Negation plays a fundamental role in language comprehension, allowing distinctions between affirmative and negative statements (e.g., ``Vocals and piano" vs. ``Vocals without piano'')~\cite{Pullum_Huddleston_Huddleston_Pullum_2002}. Existing multimodal systems struggle with encoding negation~\cite{DBLP:journals/corr/abs-2403-20312, 10.1145/3503161.3547968, 10.1145/3474085.3475241}. Also, issues with effectively representing negative statements for music applications have been raised~\cite{ismir_an_evaluation_of_vasilakis_2024, nlp4musa_music_understanding_vasilakis_2024}.

In this work, we investigate and mitigate the inability of joint audio-text models to model negation. We decide to train the models from scratch to (1) fully control the training pipeline, and (2) avoid data contamination during evaluation. We train CLAP~\cite{clap_learning_audio_concepts_elizalde_2023} models from scratch, leveraging the Million Song Dataset (MSD)~\cite{Bertin-Mahieux2011} and LP-MusicCaps-MSD captions (LP-MSD)~\cite{lp_musiccaps_doh_2023}. We experiment with two methods enabling negation modeling, using text augmentation and an additional loss term. To systematically evaluate negation modeling, we introduce novel negation evaluation protocols based on two tasks: retrieval and binary classification.


In summary, our contributions are: (1) an exploration of two, novel methods of improving negation modeling in joint audio-text models, (2) novel negation evaluation methods, framed as either retrieval or classification tasks and (3) a CLAP model trained on MSD and LP-MSD that can serve as a fair comparison point for future research, avoiding data contamination issues. All of our experiments and evaluation protocols are conducted on publicly available datasets. The code and model are publicly available for reproduction purposes.~\footnote{\href{https://www.github.com/YannisBilly/towards-effective-negation-modeling-in-joint-audio-text-models-for-music}{github.com/YannisBilly/towards-effective-negation-modeling-in-joint-audio-text-models-for-music}}


\section{Enhancing negation modeling}
\label{sec:trainig_framework}

In this section, we present the proposed approaches for enhancing negation modeling in joint audio-text models.

\subsection{Negation insert - negation text augmentation}
\label{subsec:text_augmentation}

Caption datasets are primarily synthetically generated from tagging datasets~\cite{lp_musiccaps_doh_2023}. Since negated tags (e.g. `not guitar') are largely absent from these datasets (except in MagnaTagATune~\cite{magnatagatune}), captions currently used to train audio-text models have very few instances of negation. While adding human-curated negated examples could address this, it requires a lot of resources. We propose introducing negations artificially in captions through text augmentation as a more feasible alternative.

We focus on examining the capability of models to alter their embeddings effectively in the presence of negating words (e.g. `not', `without'). We do that by randomly sampling a single music tag from the predefined tag vocabulary of the used dataset, excluding tags already present in the caption. We prepend the chosen tag with a random negating word/phrase, and place the new negated tag in a random position within the original caption. This augmentation is named \emph{Negation insert}, and will be referred to as \emph{text aug} for the remainder. We present examples of \emph{text aug} for a given original caption in Table~\ref{tab:example_of_augmentations}. Note that this approach can lead to captions with grammatical errors, but this doesn't necessarily hinder a model's retrieval performance~\cite{ismir_an_evaluation_of_vasilakis_2024}.



\begin{table}[ht]
\centering
\begin{tabular}{l p{5cm}}
  \toprule
  \textbf{Type}  & \textbf{Caption} \\
  \midrule
  Original      & A rock tune with guitar and bass \\
  Text aug 1     & A rock tune \textbf{not pop} with guitar and bass \\
  Text aug 2     & A rock tune with guitar and bass \textbf{no drums} \\
  \bottomrule

\end{tabular}
\caption{An original caption from the training dataset and its negated versions, used for \emph{Negation insert} (\emph{text aug}). The \textbf{tags} sampled for negation are unused/untrue in LP-MSD annotations.}
\label{tab:example_of_augmentations}
\end{table}

\subsection{Dissimilarity term - adding a loss term}
\label{subsec:loss_term}

It has been argued that CLIP/CLAP loss~\cite{DBLP:conf/icml/RadfordKHRGASAM21} is highly tailored to retrieval~\cite{yuksekgonul2023when, DBLP:journals/corr/abs-2403-20312}. As a result, lower-level semantic or syntactic information is not effectively encoded and utilized by the text branch~\cite{yuksekgonul2023when}. We believe that directly modeling semantic differences as an additional term in the contrastive loss will be more effective than using \emph{text aug}, which may have more subtle enhancements of negation modeling properties on joint audio-text models.

To calculate the loss term, we require negative examples in addition to the original captions. Unlike in \emph{text aug}, these captions are created by identifying tags already present in their original version. Next, we randomly prepend these tags with a negation word. Negating all tags yields \emph{Fully Negated} captions, which are used in both the proposed loss term and evaluation. We also construct \emph{Half Negated} captions by negating 50\% of the tags, used for evaluation only. Examples of both are shown in Table~\ref{tab:example_of_dissimilarity}.



Let $e_{c_{i}}$ and $ e_{\bar{c_{i}}}$ be the joint space embeddings of the caption ($c_{i}$) and its \emph{Fully Negated} version ($\bar{c_{i}}$) for the $i$-th audio-caption pair in dataset $(a_{i}, c_{i}) \in D_{train}$. We propose minimizing cosine similarity as the loss term:

\begin{equation}\label{eq:negative_term_clap_loss}
    L_\textnormal{diss} = 1 + \frac{1}{B} \cdot \sum_{i=1}^{B} (\frac{e_{c_{i}} \cdot e_{\bar{c_{i}}}}{||e_{c_{i}}||_{2} \cdot ||e_{\bar{c_{i}}}||_{2})})
\end{equation}

where $B$ is the batch size during training, $\lvert\lvert \cdot \rvert\rvert_{2}$ signifies the Euclidean norm of a vector. Therefore, the total loss is the following.

\begin{equation}\label{eq:full_loss_with_negative_term}
    L_\textnormal{total} = L_\textnormal{CLAP} + k \cdot L_\textnormal{diss} 
\end{equation}

where $k$ is the loss term weight, namely the term weight for the remainder of this paper. This method is named \emph{Dissimilarity term} and is referred to as \emph{loss term} for the remainder of this paper.

\begin{table}[ht]
\centering
\begin{tabular}{l p{5cm}}
  \toprule
  \textbf{Type}  & \textbf{Caption} \\
  \midrule
  Original      & A \textit{rock} tune with \textit{guitar} and \textit{bass} \\
  Half Negated   & A \textbf{not \textit{rock}} tune with \textit{guitar} and \textbf{without \textit{bass}} \\
  Fully Negated & A \textbf{not \textit{rock}} tune with \textbf{no \textit{guitar}} and \textbf{without \textit{bass}} \\
  \bottomrule

\end{tabular}
\caption{An original caption from the dataset and its derived negated versions. \textit{Tags already present} in the caption are \textbf{negated} for \emph{Half, Fully Negated} captions.}
\label{tab:example_of_dissimilarity}
\end{table}





\section{Evaluation of negation modeling}
\label{sec:evaluation_on_negation_modeling}

In this section, we propose two novel methodologies for evaluating and quantifying negation modeling capabilities. Both make use of the \emph{Half, Fully Negated} captions derived from the \emph{Original} ones as described in section~\ref{subsec:loss_term}.

\subsection{Negation as retrieval}
\label{subsec:negation_as_retrieval}


We hypothesize that a joint audio–text model should be sensitive to semantic negation, such that the similarity between a song and its associated captions systematically decreases as the captions are transformed from their original form to \emph{Fully Negated} versions. Therefore, we first frame negation modeling as a retrieval problem. Retrieval is performed via cosine similarity and evaluated with Recall@K (R@K), measuring how often the correct audio or caption appears within the top-K most similar results, according to cosine similarity, for text-to-audio and audio-to-text retrieval.


We expect R@K to decrease as the proportion of negated tags increases (from \emph{Original} to \emph{Fully Negated} captions). This underlines that as captions become less valid, the respective audio-caption similarity is reduced. Also, we expect R@K to approach zero for small $K$ in both \emph{Half} and \emph{Fully Negated} captions, as the correct song is likely to fall beyond the top-$K$ positions in the ranking. This motivates our choice of $K=10$ as the evaluation baseline.

\subsection{Negation as binary classification}
\label{subsec:negation_as_binary_classification}

\begin{figure}[tb]
\begin{minipage}[b]{1.0\linewidth}
  \centering
  \centerline{\includegraphics[width=8.5cm]{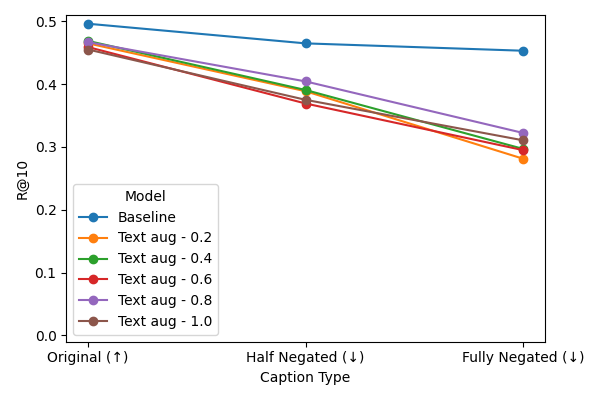}}
\end{minipage}
\caption{Sensitivity analysis of text augmentation (\emph{text aug}) using different augmentation probabilities.}
\label{fig:rat10_negation_insert}
\end{figure}

We hypothesize that joint audio–text models should encode a notion of truthfulness, such that captions more relevant to a song’s content receive higher similarity scores. To evaluate this, we frame negation modeling as a triplet-based binary classification task, where each triplet consists of an audio embedding, a more relevant caption, and a less relevant caption. After generating joint space embeddings, we compute cosine similarities between audio–relevant caption ($sim_\textnormal{a-rc}$) and audio–less relevant caption ($sim_\textnormal{a-lrc}$). Classification accuracy is defined as the proportion of cases where $sim_\textnormal{a-rc} > sim_\textnormal{a-lrc}$, noting that random choice yields an expected accuracy of 0.5.

We evaluate negation modeling by running all combinations of audio-caption pairs between \emph{Original}, \emph{Half} and \emph{Fully Negated} captions. Specifically, we compare (1) \emph{Original -- Fully}, (2) \emph{Original -- Half}, and (3) \emph{Half -- Fully}.

\section{Experiments}
\label{sec:experiments}
In this section, we present the model configurations, the dataset used and we evaluate the effectiveness of the two proposed negation modeling strategies presented in Section~\ref{sec:dataset_and_model}, both separately and combined. We first show the evaluation strategies in a retrieval setting, followed by a binary classification setting, as presented in sections~\ref{subsec:negation_as_retrieval} and~\ref{subsec:negation_as_binary_classification} respectively.

\subsection{Dataset and models}
\label{sec:dataset_and_model}
We use a CLAP~\cite{clap_learning_audio_concepts_elizalde_2023} model trained on MSD~\cite{Bertin-Mahieux2011} and LP-MSD~\cite{lp_musiccaps_doh_2023} from scratch, using the Extended Cleaned tag and Artist-Level Stratified (ECALS)~\cite{doh2023toward} subset. A CLAP model trained on LP-MSD without modifications serves as the \emph{baseline}. We then train variants with (1) \emph{text aug} augmentation probabilities from 0 to 1 and (2) a \emph{loss term} in the contrastive loss with weight $k$ from $1\mathrm{e}{-1}$ to $1\mathrm{e}{-4}$. After evaluating \emph{text aug} and \emph{loss term} separately, we test their combination. In more detail, for each audio–caption pair, we generate both a \emph{Fully Negated} caption and a single-tag negated variant (Section~\ref{subsec:text_augmentation}, Table~\ref{tab:example_of_augmentations}) and apply both methods (Sections~\ref{subsec:text_augmentation},~\ref{subsec:loss_term}). We refer to the combination as \emph{combo}. We hypothesize that mixing mildly (\emph{text aug}) and \emph{Fully Negated} (\emph{loss term}) captions helps the model capture partial semantic similarity between audio and their negated captions.

We train for up to 10 epochs and select the best checkpoint by the highest average mAP@10 across audio-to-text and text-to-audio retrieval on the LP-MSD test set. For our proposed evaluation, we randomly sample 512 songs from LP-MSD test set.

\subsection{Impact of text augmentation on retrieval metrics}
\label{subsec:evaluation_of_negation_insert}

We first test whether adding a single negated, unused tag to the \emph{Original} caption aids retrieval and/or negation modeling. Retrieval is evaluated as in Section~\ref{subsec:negation_as_retrieval} (Figure~\ref{fig:rat10_negation_insert}). The \emph{baseline} shows only a slight R@10 drop, suggesting it fails to capture the semantic change from negation. While R@10 on \emph{Original} captions remains similar across models, an augmentation probability of 0.6 yields the best and second-best R@10 on \emph{Half Negated} and \emph{Fully Negated} captions, respectively. These decrements in R@10 are minor because contrastive learning focuses on matching correct audio–caption pairs and overlooks varying degrees of dissimilarity with their negated versions. To address this, we add a loss term that explicitly separates captions from their negated counterparts.


\subsection{Impact of Dissimilarity term on retrieval metrics}
\label{subsec:evaluation_of_negation_insert}

\begin{figure}[tb]
\begin{minipage}[b]{1.0\linewidth}
  \centering
  \centerline{\includegraphics[width=8.5cm]{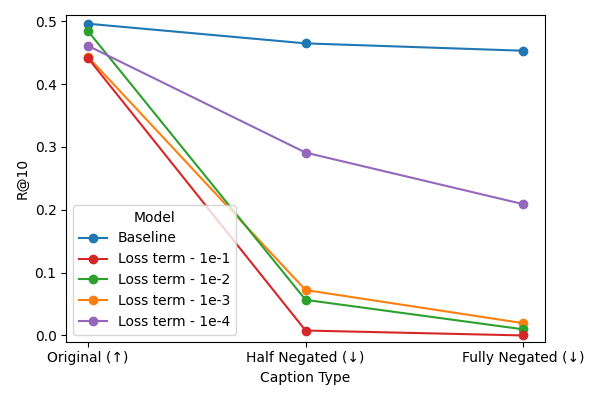}}
\end{minipage}
\caption{Sensitivity analysis of Dissimilarity term (\emph{loss term}) using different term weights $k$.}
\label{fig:rat10_dissimilarity}
\end{figure}

Next, we test whether explicitly separating \emph{Original} and \emph{Fully Negated} caption embeddings affects retrieval and/or aids negation modeling. Retrieval is evaluated as in Section~\ref{subsec:negation_as_retrieval} and the results are displayed in Figure~\ref{fig:rat10_dissimilarity}. All models trained with the \emph{loss term} show similar R@10 on \emph{Original} captions. A term weight of $1\mathrm{e}{-2}$ yields the smallest R@10 drop on \emph{Original} captions while driving R@10 to $\approx 0$ for both negated variants. These effects are stronger than using \emph{text aug}, supporting the need to explicitly separate original and negated captions in the embedding space. We proceed with combining both methods.


\subsection{Impact of the combination of proposed methods on retrieval metrics}

\begin{figure}[tb]
\begin{minipage}[b]{1.0\linewidth}
  \centering
  \centerline{\includegraphics[width=8.5cm]{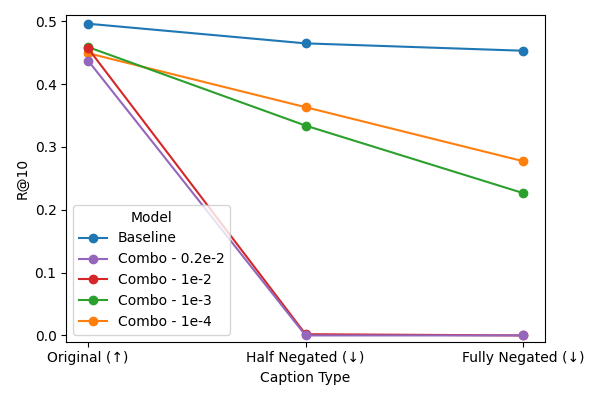}}
\end{minipage}
\caption{Sensitivity analysis of combining \emph{text aug} and \emph{loss term} (\emph{combo}) using different term weights $k$.}
\label{fig:rat10_negclap}
\end{figure}

We examine the synergy between \emph{text aug} (augmentation probability 0.6) and \emph{loss term} with different term weights $k$ and present the results in figure~\ref{fig:rat10_negclap}. Using both methods produces a more pronounced reduction in R@10 for term weights of $1\mathrm{e}{-3}$ and $1\mathrm{e}{-4}$ compared to using \emph{text aug} alone, whereas the evolution between \emph{Half} and \emph{Fully Negated} captions is more gradual than for \emph{loss term} alone.

We attribute this effect to the model perceiving negation in two ways: indirectly through audio-negated caption pairs from \emph{text aug}, which introduce a single unused negated tag with minimal semantic change, and directly through original-negated caption pairs from the \emph{loss term}, which include multiple negations. This dual exposure encourages the model to encode different degrees of semantic compatibility, capturing the distinction between \emph{Half} and \emph{Fully Negated} captions. We next evaluate the models using the proposed binary classification method (Section~\ref{subsec:negation_as_binary_classification}).

\subsection{Evaluation of negation as a binary classification task}

\begin{figure}[htb]
\begin{minipage}[b]{1.0\linewidth}
  \centering
  \centerline{\includegraphics[width=8.5cm]{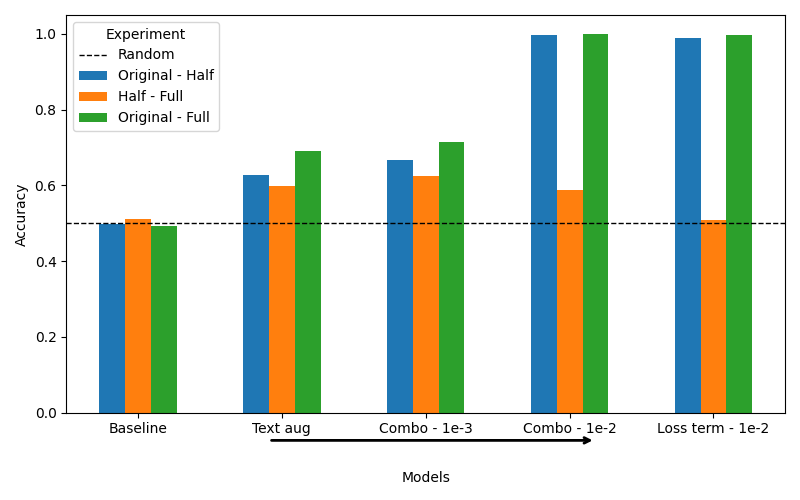}}
\end{minipage}
\caption{Binary classification evaluation of 5 CLAP variants (see section~\ref{subsec:negation_as_binary_classification}). Arrow signifies experiments where the \emph{loss term} weight increases. We highlight that \emph{text aug} is \emph{combo} with $k=0$.}
\label{fig:triplet_binary_barplot}
\end{figure}

We proceed to evaluate whether the respective CLAP variants can effectively encode that audio embeddings should be more similar to less negated versions of captions than more negated ones. We present the results in figure~\ref{fig:triplet_binary_barplot}. \emph{Baseline} achieves random choice accuracy, whereas every proposed method, independently, as well as \emph{combo}, performs better with variable success.

With the \emph{loss term} ($1e{-2}$), models learn to rank original captions above negated ones but collapse to bag-of-words behavior~\cite{DBLP:journals/corr/abs-2403-20312}, failing to distinguish \emph{Half} from \emph{Fully Negated} captions. This limitation is mitigated partially in \emph{combo}, where \emph{Half--Fully} accuracy improves beyond random while \emph{Original--Half} and \emph{Original--Fully} remain near optimal. \emph{Text aug}, alone or within \emph{combo}, is effective in binary classification, with \emph{combo} ($1e{-2}$) yielding noticeable  gains compared to \emph{baseline}, supporting their synergy. The term weight seems to introduce a trade-off: as it increases, separation of \emph{Original} to \emph{Fully Negated} captions increases, while \emph{Half--Fully} accuracy decreases. Finally and despite the lower success of \emph{text aug}, this method is essential as it helps encoding that \emph{Half Negated} are more similar to audio than \emph{Fully Negated} captions, in $\approx60\%$ of cases.

\section{Conclusion}
\label{sec:conclusion}

Our experiments show that introducing negation through text augmentation and/or an additional contrastive loss term improves the ability of audio-text models to capture negation. While these methods enhance performance on negation-specific evaluations, they lead to a slight reduction in standard retrieval performance. Also, our proposed CLAP variants only partly capture the relative similarity of \emph{Half Negated} captions and songs, which remains an area for future work.

Our methods rely on annotated tagging datasets, which often exhibit high false-negative rates~\cite{8323324} and may fail to capture the full nuance of natural negation in human-written text. This limits the reliability of negation-based evaluation, or text augmentation and, underscores the need for music tagging datasets with explicit hard negatives for controlled testing. Such datasets are essential for robust analysis of negation effects in retrieval and for enabling controlled synthetic caption generation; although their creation is resource-intensive, they represent a highly valuable direction for future work.

Lastly, while we report improvements in negation modeling, we rely on synthetic captions generated through controlled augmentation. These may not model the complexities of natural negation in human-written text, and our evaluation, limited to retrieval and binary classification, leaves open how well the model generalizes to real-world queries.

\bibliographystyle{IEEEbib}
\bibliography{strings,refs}

\end{document}